\documentclass[11pt]{article}
\usepackage{amsmath}
\usepackage{amssymb}
\usepackage{amsthm}
\usepackage{amsfonts}
\usepackage{comment}
\usepackage{color}
\usepackage{mathrsfs}
\usepackage{braket}
\usepackage{graphicx}
\usepackage[subrefformat=parens]{subcaption}
\usepackage{cite}
\usepackage{here}
\usepackage{tikz}
\usetikzlibrary{decorations.pathmorphing}
\usepackage{caption}

\usepackage[stable]{footmisc}

\makeatletter

\@addtoreset{equation}{section}
\makeatother

\begin{document}

\pagestyle{empty}

\title{\bf{Effects of fluctuations\\in higher-dimensional AdS black holes}}

\date{}
\maketitle
\begin{center}
{\large 
Hyewon Han\footnote{dwhw101@dgu.ac.kr}, Bogeun Gwak\footnote{rasenis@dgu.ac.kr}
} \\
\vspace*{0.5cm}

{\it 
Division of Physics and Semiconductor Science, Dongguk University, Seoul 04620,\\ Republic of Korea
}

\end{center}

\vspace*{1.0cm}
\begin{abstract}
{\noindent
We explored the impact of mass fluctuations on anti-de Sitter black holes in higher dimensions, particularly focusing on their effects on thermodynamic properties and null trajectories of the black holes. Our findings indicate that mass oscillations lead to perturbations in thermodynamic variables and geodesics. These result in the second-order fluctuations for the location of the horizon, thereby altering the Hawking temperature and Bekenstein--Hawking entropy. Furthermore, we derived equations for perturbed null rays near the horizon with arbitrary dimensions and for complete null rays in the large $D$ limit. 
}
\end{abstract}

\newpage
\baselineskip=18pt
\setcounter{page}{2}
\pagestyle{plain}
\baselineskip=18pt
\pagestyle{plain}
\setcounter{footnote}{0}

\section{Introduction}
A black hole, in classical gravity theory, is an entity that completely absorbs any matter falling onto its surface. However, accounting for quantum mechanical effects in curved spacetime reveals that black hole emits thermal radiation, a mechanism demonstrated by Hawking \cite{hawking1975particle}. Quantum fluctuations governed by the uncertainty principle spontaneously generate energy quanta near the event horizon of a black hole, and a distant observer measures the thermal spectrum. The black hole can be in thermal equilibrium with its radiation and is specified by the Hawking temperature, which is proportional to its surface gravity. It was also proposed that a black hole has Hawking--Bekenstein entropy proportional to the area of the horizon \cite{bekenstein2020black,bekenstein1973black,bekenstein1974generalized}. Furthermore, these black hole features, which are the result of considering quantum effects, resemble the laws of thermodynamics \cite{bardeen1973four}. The thermodynamic nature of black holes have been a focus of recent extensive research.

Using a semi-classical treatment of quantum fluctuations, York \cite{york1983dynamical} presented a dynamic description of the Hawking effect. This model involves a spherical, asymptotically flat black hole oscillating in its gravitational quasi-normal mode. It calculates the locations of the black hole’s horizons for both spherical and non-spherical oscillations, and the results suggest the formation of a `quantum ergosphere' through which matter can tunnel out of the black hole. Building upon this work, Barrab$\grave{\mathrm{e}}$s et al. \cite{barrabes1999metric} modified the Hawking radiation caused by fluctuations in the black hole geometry. They also examined the propagation of a massless field by considering a stochastic ensemble of fluctuations \cite{barrabes2000stochastically}. These stochastic fluctuations arising from quantum matter fields, can be described within the framework of stochastic semi-classical gravity \cite{hu2007metric,hu2008stochastic}. Furthermore, the fluctuating black hole model was extended to higher-dimensional gravity, and the dimensional dependencies of the corrections due to spherical oscillations were explored in \cite{han2023metric}. Various aspects of fluctuating black hole geometries are also addressed in different contexts in \cite{ford1997cosmological,parentani2001quantum,bellucci2010thermodynamic,arias2012thermal,brustein2013restoring,iofa2016density,frolov2017quantum}.

In this context, we are considering a gravity theory with a negative cosmological constant. The anti-de Sitter (AdS) spacetime, a maximally symmetric solution of the Einstein field equations, has a constant negative curvature and a boundary at spatial infinity. In particular, AdS spacetime is crucial in conformal field theory (AdS/CFT) correspondence \cite{maldacena1999large,gubser1998gauge,MR1633012,aharony2000large}, which postulates the relationship between gravity theory on AdS spacetime and CFT defined on the AdS boundary at zero temperature. Namely, $(D-1)$-dimensional quantum field theory can be described using $D$-dimensional gravity theory. CFT at finite temperatures is dual to the physics of black holes in AdS spacetime \cite{MR1633012,MR1646895} because black holes behave as thermodynamic systems. In this context, AdS/CFT correspondence has prompted active research into AdS black holes in arbitrary dimensions \cite{Gwak:2018akg,chabab2020thermodynamic,garbiso2020hydrodynamics,mansoori2020universal,Gwak:2021tcl,aref2021holographic,wu2021hawking,Kan:2021blg,gwak2022violation,liu2023topological}.

Given that gravity theories in higher dimensions typically entail complex and nonlinear dynamics, it is crucial to consider scenarios where the number of dimensions, $D$, approaches infinity \cite{emparan2013large,emparan2013large2,emparan2020large}. In the context of high-dimensional spacetime ($D$), the radial gradient of the gravitational potential near the event horizon of a black hole becomes exceedingly steep. Beyond the event horizon, in the exterior region of a narrow radial span of approximately $1/D$, gravitational effects become negligible. This corresponding length scale allows us to effectively separate the vicinity of the horizon from distant regions. Near the horizon, the gravitational influence of the black hole is predominant, while in the outer region, the black hole behaves much like a non-interacting particle within a flat (in our case, AdS) spacetime. This limit fundamentally transforms the dynamics, leading to an effective theory on the surface of the black hole’s horizon. The $1/D$ expansion simplifies complex calculations significantly and facilitates an analytical approach in various recent studies. It has been applied to investigate a wide range of topics, including the spectra of quasi-normal modes and the stability of static black holes\cite{emparan2014universal,emparan2014decoupling}, rotating black holes \cite{emparan2014instability,suzuki2015stationary,tanabe2016black,chen2017charged}, (A)dS black holes \cite{emparan2015quasinormal,tanabe2016instability,herzog2018large}, and black string and brane solutions \cite{emparan2015evolution,suzuki2015non,rozali2016brane,emparan2018phases}. Furthermore, the large $D$ limit has also been employed to explore the dynamics of solutions in the Einstein--Gauss--Bonnet theory \cite{chen2016quasinormal,chen2017static,chen2017einstein,suzuki2023phase}. 

In this work, we investigated mass fluctuations in higher-dimensional AdS black holes. Incorporating a negative cosmological constant, our model represents a direct generalization of the fluctuating black hole discussed in \cite{york1983dynamical,barrabes1999metric,han2023metric}. Given the pivotal role that mass plays in the context of black hole spacetime, its fluctuations have a profound impact, not only on the geometry but also on the thermodynamics and the trajectories of geodesics. Mass fluctuation was introduced as a small oscillation in mass, which manifests as its effects prominently in the metric. Solving the equation of motion reveals an oscillation in the black hole's horizon, which we have determined up to the second order. In particular, the oscillatory horizon leads to a change in key thermodynamic variables such as the Hawking temperature and Bekenstein--Hawking entropy. This allows us to gain insights into the preferred state of the black hole within the thermodynamic analysis. Furthermore, the oscillations in mass perturb geodesics within the spacetime, and we have derived a general solution for the propagation of null rays in close proximity to the horizon. Determining the complete solution for an arbitrary location in the geodesic equation becomes considerably intricate in AdS spacetime due to integrals in general dimensions. However, focusing on the gravitational influence's strength enables us to render the geodesic equation solvable by employing a higher-dimensional limit called the large $D$ limit. Our comprehensive solution for the large $D$ limit demonstrates the perturbed trajectory of the null ray.

The remainder of this paper is organized as follows: In Section 2, we review the fluctuating black hole model and generalize it to AdS spacetime. In Section 3, we present the perturbed position of an event horizon and calculate the thermodynamic variables. In Section 4, we derive a solution describing the trajectory of a null ray propagating in a fluctuating black hole geometry. In Section 5, we examine the solution to a large $D$ limit. The results are summarized in Section 6. We employed a metric signature $(-,+,+,+, \cdots)$ and its units, where $G_D=c=1$.

\section{Model of fluctuating black holes}
The influence of metric fluctuations has been investigated for four \cite{barrabes1999metric} and higher dimensions \cite{han2023metric}. To classically treat quantum fluctuations near the event horizon of a black hole, they employed the fluctuating black hole model proposed in \cite{york1983dynamical}. This is described by an ingoing Vaidya-type solution with an oscillating mass. For $D\ge4$, the metric is given by \cite{iyer1989vaidya}
        \begin{align} \label{metric}
            ds^2 = - \left( 1- \frac{m}{r^{D-3}} \right) d v^2 + 2 dv dr + r^2 d \Omega_{D-2}^2,
        \end{align}
where $d \Omega_{D-2}^2$ are the line elements on the unit $(D-2)$ spheres. The mass function is given by
        \begin{align} \label{masss}
            m = m(v,\theta) = M \left[ 1 + \sum_l (2l+1) \mu_l \sin (\omega_l v) Y_l(\theta) \right] \vartheta (v),
        \end{align}
where $M$ is a mass parameter related to the black hole mass.
        \begin{align}
            M_{BH}=\frac{(D-2)\Omega_{D-2}}{16 \pi } M,
        \end{align}
and $\Omega_{D-2}=\frac{2 \pi^{(D-1)/2}}{\Gamma(\frac{D-1}{2})}$ denotes the area of the unit $(D-2)$ sphere. The second term in the brackets of \eqref{masss} represents the fluctuating part given by the sum of the resonant modes for each angular index $l$. The dimensionless amplitude parameter $\mu_l=\alpha_l \hbar^{1/2} M_{BH}^{-1}$, where $\alpha_l$ is a pure number, is extremely small for massive black holes. $\omega_l$ indicates the resonant frequencies, and $Y_l(\theta)$ are the normalized spherical harmonics on unit $(D-2)$ spheres with a zero azimuthal index. The angular coordinates, except for the azimuthal angle, are collectively denoted as $\theta=(\theta_1, \cdots, \theta_{D-3})$. The Heaviside step function $\vartheta(v)$ is included assuming that a black hole is formed by the gravitational collapse of a spherical null shell with mass $M$. Therefore, metric \eqref{metric} describes the flat spacetime inside the null shell when $v<0$ and the fluctuating black hole spacetime when $v>0$. 
If we consider the lowest mode $l=0$ as a simple case, we have $Y_0(\theta)=1$ \cite{zhao2017spherical,erdelyi1953higher} and
        \begin{align} \label{mass}
            m(v)=M\left[1+\mu_0 \sin (\omega v) \right] \vartheta(v),
        \end{align}
where the subscript $0$ of the frequency is omitted. It describes spherical oscillations with a small amplitude $\mu_0$ and a period $2\pi/\omega$. The study focused on examining the effects of fluctuations on the event horizon and outgoing rays in black holes. This was achieved by solving a null geodesic equation specific to this geometry. The research aimed to deepen understanding of black hole dynamics under the influence of perturbations.

We generalize this model to a theory with the cosmological constant $\Lambda$. The Einstein field equations are obtained from the following actions in $D$ dimensions.
        \begin{align} \label{action}
            I = \frac{1}{16\pi} \int d^D x \sqrt{-g} \, (R-2\Lambda) + I_{matter}. 
        \end{align}
The cosmological constant is given by
        \begin{align}
            \Lambda = \pm \frac{(D-1)(D-2)}{2 L^2},
        \end{align}
where $L$ is the curvature radius of the de Sitter (dS) or anti-de Sitter (AdS) spacetime; the positive sign corresponds to dS, and the negative sign corresponds to AdS. The action principle yields the following field equations.
        \begin{align}
            G_{\mu \nu} + \Lambda g_{\mu\nu} = 8 \pi  T_{\mu \nu},
        \end{align}
where $T_{\mu \nu}$ is the stress-energy tensor. We consider a spherically symmetric solution with negative $\Lambda$. A fluctuating black hole geometry can be approximated using a higher-dimensional Vaidya-AdS metric. This is written in the ingoing Eddington--Finkelstein coordinates $(v,r)$ as \cite{ghosh2008radiating}
        \begin{align}
            ds^2 = - f(v,r) d v^2 + 2 d v d r + r^2 d \Omega^2_{D-2},
        \end{align}
where
        \begin{align}
            f(v,r) = 1 - \frac{m(v)}{r^{D-3}}+ \frac{r^2}{L^2}.
        \end{align}
To simplify the model, we consider only the spherically oscillating mass \eqref{mass}. Subsequently, we obtain the stress-energy tensor in the form
        \begin{align}
            T_{\mu \nu} = \frac{D-2}{16 \pi r^{D-2}} \left[ M (1+\mu_0 \sin ( \omega v)) \delta(v) + M\mu_0 \omega \cos(\omega v) \vartheta(v) \right] l_{\mu} l_{\nu},
        \end{align}
where $l_{\mu}=-\partial_{\mu} v$ is a null vector field tangent to the incoming null geodesics. The first term, containing the delta function, corresponds to a null shell with mass $M$ and additional fluctuating and ingoing null dust. The second term indicates that the energy density in the black hole spacetime fluctuates around a mean of zero.

The causal structure of this spacetime without fluctuations, that is, $\mu_0=0$, is shown in Figure \ref{fig:diagram}. The vertical line on the right side of the diagram represents the AdS boundary $r=\infty$ and the line on the left side represents a regular origin $r=0$. The doubled line is a null shell propagating along $v=0$, and the dashed line after collapse represents the future event horizon $r=r_H$ of the black hole. In the study, it was observed that for rays to propagate outward in black hole spacetime without entering the trapped region, incoming rays leaving the boundary must depart earlier than the trajectory represented by the dashed line. The focus was on observing outgoing rays reaching the boundary, disregarding those reflected at the boundary and traveling towards the singularity. Metric fluctuations were found to modify the trajectories of rays in black hole geometry, highlighting the intricate dynamics influenced by these perturbations.

\begin{center}
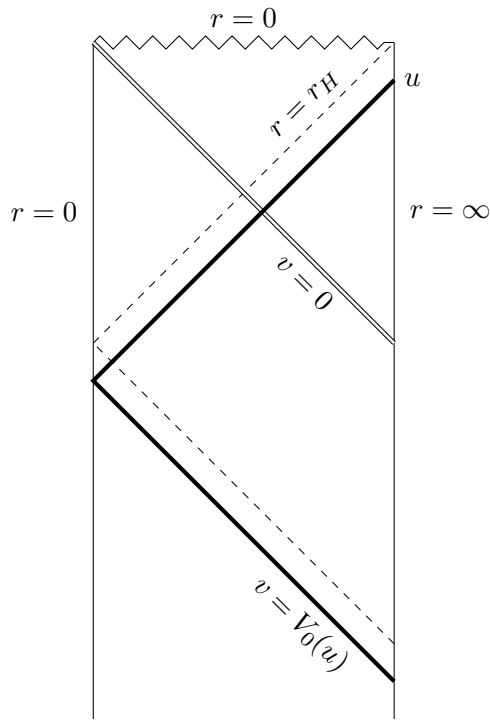

\begin{tikzpicture} 

    \node (I)    at (4,0)  {};
    \node (a)    at (4,-2) {};

    \path
     (I) +(90:4)  coordinate (Itop)
         +(-90:4) coordinate (Ibot)
         +(-90:5) coordinate (bot)
         +(180:4) coordinate (Ileft)
       ;
    \draw[dashed] (Ileft) -- node[near end, above, sloped] {$r=r_H$}
                  (Itop);
    \draw (Itop) --      
          (bot) node[near start, right, inner sep=2mm]    {$r=\infty$} ;
    \draw[dashed]
          (Ibot) --  (Ileft);
      
    \path 
     (Itop) + (-4,0) coordinate (sing);
    \path 
      (sing) +(0,-9) coordinate (past);
    \draw (sing) -- (past) node[near start, left, inner sep=2mm]    {$r=0$};
    \draw[decorate,decoration=zigzag] (sing) -- (Itop)
         node[midway, above, inner sep=2mm] {$r=0$};

    \path
      (I) +(0,0) coordinate (v0);
    \draw[thin,double distance=1pt]
        (v0) -- 
          node[near start, below, sloped] {$v=0$}
        (sing);

    \path
      (Ileft) +(0,-0.5) coordinate (bounce);
    \path
     (Ibot)  +(0,-0.5) coordinate (Vu);
    \path
     (Itop) +(0,-0.5) coordinate[label=0:$u$] (bounce2);
    \path
     (Itop) +(-0.5,0) coordinate (end);
    \draw[ultra thick]
     (Vu) -- node[near start, below, sloped] {$v=V_0(u)$} (bounce) -- (bounce2)  ;
    
\end{tikzpicture} 
\captionof{figure}{Conformal diagram of the asymptotically AdS black hole formed by a gravitational collapse of a null shell without fluctuations. The collapsing null shell propagated along the double line $v=0$. The zigzag line indicates the singularity of the black hole, and the thick solid line represents the trajectory of a radial null ray propagating near an event horizon.}
\label{fig:diagram}
\end{center}

To examine the influence of fluctuations on the event horizon and propagating rays, we solved a null geodesic equation for radially outgoing rays in the region $v>0$.
        \begin{align} \label{geodesic}
            1 - \frac{M\left[1+\mu_0 \sin (\omega v) \right]}{r^{D-3}}+ \frac{r^2}{L^2} = 2 \frac{d r }{d v}.
        \end{align}
As the amplitude parameter $\mu_0$ of the fluctuations is assumed to be minuscule, we can employ a perturbation method. We set the radial coordinates to
        \begin{align} \label{perturbation}
            r = r (v) = R (v) + \rho (v) + \sigma (v) + \cdots,
        \end{align}
where $R(v)$ is a solution without fluctuations and $\rho(v)$ and $\sigma(v)$ correspond to the first- and second-order perturbations in $\mu_0$, respectively. The higher-order terms are denoted by dots, and we work up to the second order. In the following calculations, we fix the value of the retarded time coordinates as $u=v-2R_*$, where
        \begin{align} \label{tortoise}
            R_*=\begin{cases}
L \arctan \left(\frac{R}{L}\right), & \mbox{for } v<0 \\
\int \left(1 - \frac{M}{R^{D-3}}+ \frac{R^2}{L^2}\right)^{-1} d R, & \mbox{for }v>0
\end{cases}
        \end{align}
which is the radial tortoise coordinate without fluctuations. This requirement corresponds to a condition in which the rays propagating outward in the perturbed black hole geometry reach the boundary at the same retarded time as the rays propagating in the unperturbed geometry. Subsequently, $u$ is a fixed parameter specifying the unperturbed trajectory $r=R(v;u)$ of the ray.

By substituting \eqref{perturbation} into the geodesic equation \eqref{geodesic} for $v>0$ and linearizing, we obtain         
        \begin{align}
            2 &\, \frac{d R }{d v} = 1 - \frac{M}{R^{D-3} } +\frac{R^2}{L^2}  , \label{zeroth} \\
            2 &\, \frac{d \rho }{d v} - \left[(D-3) \frac{M}{R^{D-2} } +\frac{2 R}{L^2}\right] \rho = - \frac{M}{R^{D-3} } \mu , \label{firstpe} \\ 
            2 &\, \frac{d \sigma }{d v} - \left[(D-3) \frac{M}{R^{D-2} }+\frac{2 R}{L^2}\right] \sigma = (D-3) \frac{M}{R^{D-3} } \left[ \frac{\rho }{R}  \mu - \frac{D-2}{2} \frac{\rho^2}{R^2}  \right]+\frac{\rho^2}{L^2}, \label{secondpe}
        \end{align}
where $\mu=\mu_0 \sin (\omega v)$. It is convenient to write the perturbation equations \eqref{firstpe} and \eqref{secondpe} in common their form as 
        \begin{align} \label{common}
            \frac{d k}{d v} - \left[\frac{D-3}{2} \frac{M}{R^{D-2}} +\frac{R}{L^2} \right] k = K,
        \end{align}
where the first-order perturbation corresponds to
        \begin{align} \label{eq:first2}
            k = \rho, \qquad K = -  \frac{M}{2R^{D-3} }  \mu.
        \end{align}
The second-order perturbation corresponds to
        \begin{align} \label{eq:second2}
            k = \sigma, \qquad K = \frac{D-3}{2} \frac{M}{R^{D-3} } \left[ \frac{\rho}{R} \mu - \frac{D-2}{2} \frac{\rho^2}{R^2} \right] + \frac{\rho^2}{2 L^2}.
        \end{align}
In the following chapters, we solve these perturbation equations to seek solutions describing the perturbed position of an event horizon and the perturbed trajectories of the rays propagating in black hole geometry.

\section{Perturbed event horizons and thermodynamics}
We begin by considering an unperturbed solution $R=R_H$ and determine a perturbed solution $r=r_H=R_H+\rho_H+\sigma_H+\cdots$ for $v>0$, describing the position of the event horizon of a fluctuating black hole. The unperturbed position $R_H$ satisfies
        \begin{align}
            R_H^{D-3} \left( 1+ \frac{R_H^2}{L^2} \right) - M = 0.
        \end{align}
The unperturbed surface gravity is
        \begin{align}
            \kappa = \frac{D-3}{2} \frac{M}{R_H^{D-2}} + \frac{R_H}{L^2}.
        \end{align}
Using these variables, the perturbation equations are expressed as
        \begin{align}
            \frac{d \rho_H}{d v} - \kappa \rho_H =& - \frac{\mu }{2} \left( 1+ \frac{R_H^2}{L^2} \right), \\
            \frac{d \sigma_H}{d v} - \kappa \sigma_H =& \frac{D-3}{2}\left( 1+ \frac{R_H^2}{L^2} \right) \left[ \frac{\rho_H}{R_H}\mu - \frac{D-2}{2} \frac{\rho_H^2}{R_H^2} \right] + \frac{\rho_H^2}{2 L^2}.
        \end{align}
By solving these first-order differential equations, we obtained the perturbed position of the event horizon.
        \begin{align}
            \rho_H = & \frac{\mu_0}{2 \kappa} \left( 1+ \frac{R_H^2}{L^2} \right) \frac{ \Omega \cos (\omega v) + \sin (\omega v)}{1+ \Omega^2} ,\\
            \sigma_H =& \frac{\mu_0^2}{4 \kappa^2} \left( 1+ \frac{R_H^2}{L^2} \right)^2 \left[ \frac{(D-3)}{2 R_H} \frac{ 2\Omega^2 (2 - \Omega^2) \cos (2\omega v) + \Omega(1-5\Omega^2) \sin (2 \omega v)}{(1+ \Omega^2)^2 (1+4\Omega^2)} \right. \nonumber \\
            & \left. - \frac{1}{4 \kappa} \left\{ \frac{(D-3)(D-4)}{ R_H^2} +\frac{(D-1)(D-2)}{L^2} \right\} \right. \nonumber \\
            & \left. \times \frac{ (1-5\Omega^2) \sin^2 (\omega v) + \Omega (2-\Omega^2) \sin (2 \omega v) +   \Omega^2(5+2\Omega^2) }{(1+ \Omega^2)^2 (1+4\Omega^2)} \right],
        \end{align}
where $\Omega=\omega/\kappa$ denotes a dimensionless frequency. Here, the integration constants are chosen to eliminate terms that cause fluctuations $\rho_H(v)$ and $\sigma_H(v)$ to increase exponentially over time $v$. These results indicate that the event horizon fluctuates with the amplitudes and frequencies determined by the metric fluctuation parameters.

Small and periodic changes in the position of the event horizon affect the thermodynamic variables. We calculated their mean values by averaging over time $v$ to investigate the overall changes and denote them with an overbar. The mean values of the surface area and surface gravity of the perturbed black hole geometry were calculated as follows.
        \begin{align} 
            \overline{\mathcal{A}} &\equiv \Omega_{D-2}\, \overline{ (r_H^{D-2}(v))} \nonumber \\ 
            & = \Omega_{D-2} \, R_H^{D-2} \left[1+\frac{\mu_0^2}{4 (1+\Omega^2)} \left( 1+ \frac{R_H^2}{L^2} \right)^2 \frac{(D-2)\left\{(D-3)-(D-1) R_H^2/L^2 \right\}}{\left\{(D-3)+(D-1) R_H^2/L^2 \right\}^3} \right],  \label{area} \\
            \overline{\kappa} &\equiv \frac{D-3}{2}M \overline{\left( \frac{1+\mu_0 \sin (\omega v)}{r_H^{D-2}(v)} \right)} + \frac{\overline{(r_H(v))}}{L^2} \nonumber \\ 
            &= \kappa \left[1+\frac{\mu_0^2}{4 (1+\Omega^2)}\left( 1+ \frac{R_H^2}{L^2} \right)^2 \right. \nonumber \\
            & \left. \qquad \times \frac{(D-2)(D-3)^2+(D-1) R_H^2/L^2\left\{4(D-3)-(D-1)(D-2)R_H^2/L^2 \right\} }{\left\{(D-3)+(D-1) R_H^2/L^2 \right\}^4} \right]. \label{surgra}
        \end{align}
The mean value of the Hawking temperature can be obtained from the relation $\overline{T}_H = \hbar \, \overline{\kappa} /(2 \pi k_B)$, where $k_B$ is the Boltzmann constant. Subsequently, the changes in the surface area $\delta \mathcal{A} = \overline{\mathcal{A}} - \mathcal{A}$ and Hawking temperature $\delta T_H = \overline{T}_H - T_H$ caused by fluctuations are related by
        \begin{align}
            \frac{\delta \mathcal{A}}{\mathcal{A}} = \frac{\delta T_H}{T_H} - \frac{\mu_0^2}{1+\Omega^2} \left( 1+ \frac{R_H^2}{L^2} \right)^2 \frac{(D-1)(D-3)R_H^2/L^2}{\left\{(D-3)+(D-1)R_H^2/L^2\right\}^4},
        \end{align}
where $\mathcal{A}$ and $T_H$ are values without fluctuations. Unlike the asymptotically flat case in \cite{han2023metric}, the fluctuating AdS black hole features a relation that includes an additional term that is dependent on the dimension number and curvature radius. The entropy of the black hole also fluctuates. By identifying the energy with the mean mass of the fluctuating black hole $E=\overline{m(v)}=M$ and using the first law $dE = \overline{T}_H d\overline{S}$, the mean value of the entropy is calculated as follows.
        \begin{align} \label{entro}
            \overline{S} \simeq & \frac{k_B \overline{\mathcal{A}}}{4 \hbar} \left[1- \frac{\mu_0^2}{2 (1+\Omega^2)} \left( 1+ \frac{R_H^2}{L^2} \right)^2 \right. \nonumber \\
            & \left. \times \frac{(D-2)(D-3)^2 + (D-1)R_H^2/L^2 \left\{2(D-3)-(D-1)(D-2)R_H^2/L^2\right\}}{\left\{(D-3)+(D-1)R_H^2/L^2\right\}^4} \right].
        \end{align}
This indicates that the standard relationship between the entropy and surface area was modified. The value is slightly less than the entropy of the stationary spacetime without fluctuations, $\mu_0=0$. Because a system with higher entropy is preferred, this implies that after an extended period, the black hole absorbs a fluctuating null flux, and the system can evolve into a stationary spacetime according to the second law $\delta S \ge 0$. Moreover, if the other parameters $L$ and $R_H$ are fixed, the correction term decreases as the number $D$ of spacetime dimensions increase. We recover the results \cite{han2023metric} for the asymptotically flat case when $1/L^2=0$.

\section{Propagating null rays in fluctuating geometry}

We will now calculate the general solutions for $R > R_H $ and $v>0$. This pertains to the trajectory of a perturbed ray propagating outward beyond an event horizon. To achieve this, we employ the zeroth-order equation \eqref{zeroth} and substitute the advanced time parameter $v$ in the perturbation equation \eqref{common} with $R(v;u)$, where $u$ is the fixed retarded time parameter: Subsequently, we have
        \begin{align} \label{vc}
            \left(1-\frac{M}{R^{D-3}}+\frac{R^2}{L^2} \right) \frac{d k}{d R}-\left[ (D-3)\frac{M}{R^{D-2}}+\frac{2R}{L^2} \right] k = 2K.
        \end{align}
By solving this equation, we obtain
        \begin{align}
            k = \left(1-\frac{M}{R^{D-3}}+\frac{R^2}{L^2}\right) \left[-\int^{\infty}_{R}\frac{2 K}{\left(1-\frac{M}{R'^{D-3}}+\frac{R'^2}{L^2}\right)^2} \, d R' + k_{\infty} \right],
        \end{align}
where 
        \begin{align}
            k_{\infty} = \lim_{R \to \infty} \frac{k(R)}{\left(1-\frac{M}{R^{D-3}}+\frac{R^2}{L^2}\right)},
        \end{align}
which denotes the integration constant. In order for the perturbed rays to reach the boundary in the same retarded time as the unperturbed rays, we make the assumption that $k_{\infty}=0$ \cite{barrabes1999metric}.
Therefore, we have
        \begin{align} 
            \rho&=\left( 1-\frac{M}{R^{D-3}}+\frac{R^2}{L^2} \right) \int^{\infty}_{R} \left(1-\frac{M}{R'^{D-3}}+\frac{R'^2}{L^2}\right)^{-2} \frac{ M}{R'^{D-3}} \mu \, d R', \label{ff} \\
            \sigma&=-\left( 1-\frac{M}{R^{D-3}}+\frac{R^2}{L^2} \right) \int^{\infty}_{R} \left(1-\frac{M}{R'^{D-3}}+\frac{R'^2}{L^2}\right)^{-2} \nonumber \\
            &  \qquad \qquad \qquad \qquad \qquad \qquad  \quad \times \left\{(D-3)\frac{ M}{R'^{D-3}}\left(\frac{\rho}{R'} \mu- \frac{D-2}{2} \frac{\rho^2}{R'^2} \right) +\frac{\rho^2}{L^2} \right\} \, d R'. \label{ss}
        \end{align}
The fluctuating part of the mass function is written as
        \begin{align} \label{flucmass}
            \mu=\mu(R')=\mu_0 \sin \left[ \Omega(\tilde{u}+2\kappa R_*(R')) \right],
        \end{align}
where $R_*$ denotes the tortoise coordinates in \eqref{tortoise} for $v>0$, and $\tilde{u}=\kappa u$.

We consider a ray reaching the boundary at later time $u$. Because such a ray propagates near the event horizon in black hole geometry, its effective frequency becomes extremely high. Thus, we can track it backward in time using the geometric optics approximation. The trajectory of the ray without fluctuations, that is, $\mu_0=0$, is indicated by the thick solid line in Figure. \ref{fig:diagram}. In the region $v>0$, the ray propagates close to the event horizon and crosses the null shell at $v=0$. It travels in AdS spacetime inside the shell $v<0$, bounces off at the regular origin $r=R=0$, and diverges toward the boundary. One can establish a relation $V_0(u)$ between the value of the advanced time $v$ when the ray leaves the boundary and the value of the retarded time $u$ when it arrives at the boundary. When $v<0$, we obtain the following relationship.
        \begin{align} \label{V0}
            V_0=-2L \arctan \left( \frac{R_0}{L} \right),
        \end{align}
where $R_0$ denotes the value of the unperturbed radial coordinate $R$ on the null shell $v=0$. When $v>0$, we have
        \begin{align} \label{eq:V01}
            u = - 2 \int \left(1 - \frac{M}{R_0^{D-3}}+ \frac{R_0^2}{L^2}\right)^{-1} d R_0.
        \end{align}
Because the value of the radial coordinate of the ray should be continuous across the shell $v=0$, we obtain the relation $V_0(u)$ by combining the equations above. Fluctuations affect the trajectory of the ray in the region $v>0$ and thus modify the $V_0(u)$ function. In the presence of fluctuations $\mu_0 \ne 0$, Equation \eqref{eq:V01} remains unchanged, whereas Equation \eqref{V0} is modified to
        \begin{align} \label{V}
            V=-2L \arctan \left[ \frac{R_0 + \rho(R_0) + \sigma(R_0)}{L}  \right].
        \end{align}
By calculating Equations \eqref{ff}, \eqref{ss}, and \eqref{eq:V01} for $v=0$ and large $u$, we obtain the $V(u)$ function from relation \eqref{V}, which illustrates the effect of metric fluctuations in the black hole geometry.

\section{The ray propagation in the large $D$ limit}
The complex nature of the fluctuating component \eqref{flucmass} of the mass function, which includes the tortoise coordinates, poses a challenge for analytical solutions in general dimensions. In this section, we demonstrate how simplifications can be applied by taking advantage of a large $D$ limit \cite{emparan2013large}, enabling the derivation of a comprehensive solution.

In the scenario of a significantly high number of spacetime dimensions, where $D \gg 1$, the radial gradient of the gravitational potential becomes extremely steep. Consequently, the gravitational field becomes highly localized within a narrow radial range near the event horizon of a black hole. In this context, we can examine the black hole’s geometry by segregating the regions very close to the horizon from those farther away. To investigate a ray reaching the boundary at a later time $u$ in the `near-horizon zone,' we introduce a new coordinate 
        \begin{align}
            \hat{\mathsf{R}} \equiv \left( \frac{R}{R_H} \right)^{D-3},
        \end{align}
defined by $\ln \hat{\mathsf{R}} \ll D-3$. Using this near-horizon coordinate and considering up to the first order in $1/D$, the unperturbed tortoise coordinate for $v>0$ is calculated as
        \begin{align}
            d R_* = \left(1 - \frac{M}{R^{D-3}}+ \frac{R^2}{L^2}\right)^{-1} \, d R \, \simeq \, \frac{R_H}{D (1+R_H^2/L^2)} \frac{1}{\hat{\mathsf{R}}-1} \, d \hat{\mathsf{R}},
        \end{align}
yielding
        \begin{align}
            R_* = \frac{R_H}{D(1+R_H^2/L^2)} \ln (\hat{\mathsf{R}}-1).
        \end{align}
Note that $(D-3)$ becomes $D$ because we take $D$ to be very large. Subsequently, we have the fluctuating part of the mass function in the form 
        \begin{align}
            \mu (\hat{\mathsf{R}} ) &= \mu_0 \sin \left[ \Omega \left\{ \tilde{u} + \left(1+\frac{2R_H^2}{D(R^2_H+L^2)} \right) \ln (\hat{\mathsf{R}}-1) \right\} \right] \nonumber \\
            &= \mu_0 \mathrm{I m} \left[ e^{i \Omega \tilde{u}} (\hat{\mathsf{R}}-1)^{i \Omega \left(1+ \frac{2R_H^2}{D(R^2_H+L^2)} \right)} \right].
        \end{align}
The first-order perturbation is as follows
        \begin{align} \label{largefp}
            \rho (\hat{\mathsf{R}}) =\frac{R_H}{D} \left(1-\frac{1}{\hat{\mathsf{R}}}\right) I(\hat{\mathsf{R}}),
        \end{align}
where
        \begin{align}
            I(\hat{\mathsf{R}}) &= \int^{\infty}_{\hat{\mathsf{R}}} \frac{\mu(\tau)}{\left(\tau - 1\right)^2}\,d \tau =  \mu_0 \mathrm{I m} \left[e^{i \Omega \tilde{u}} \int^{\infty}_{\hat{\mathsf{R}}} \left(\tau - 1\right)^{-2+i\Omega \left(1+\frac{2R_H^2}{D(R^2_H+L^2)} \right)}\,d \tau \right] \nonumber\\ 
            &= \mu_0 \mathrm{I m} \left[ e^{i \Omega \tilde{u}} \frac{1 + i\Omega}{1+\Omega^2} ( \hat{\mathsf{R}} -1 )^{-1+i \Omega\left(1+\frac{2R_H^2}{D(R^2_H+L^2)} \right)} \right].
        \end{align}
To calculate the $V(u)$ function, we need a value on the null shell $v=0$ that satisfies
        \begin{align} \label{shellcon}
            \tilde{u} + \left(1+\frac{2R_H^2}{D(R^2_H+L^2)} \right) \ln (\hat{\mathsf{R}}_0-1 ) = 0,
        \end{align}
where $\hat{\mathsf{R}}_0$ denotes the value of $\hat{\mathsf{R}}$ at $v=0$. By imposing this condition, we obtain the following solution on the shell.
        \begin{align} \label{result1}
            \rho(\hat{\mathsf{R}}_0) &\simeq \frac{\mu_0 R_H \Omega }{D(1+\Omega^2)\hat{\mathsf{R}}_0}.
        \end{align}
Next, the second-order perturbation is rewritten in near-horizon coordinates as follows.
        \begin{align}
            \sigma (\hat{\mathsf{R}}) = - \left(1-\frac{1}{\hat{\mathsf{R}}}\right) \int^{\infty}_{\hat{\mathsf{R}}} \frac{\rho(\tau)}{(\tau - 1)^2}   \left\{ \mu(\tau) - \frac{D}{2R_H}\rho(\tau) \right\}\,d \tau. 
        \end{align}
Using Equation \eqref{largefp}, we obtain
        \begin{align}
            \sigma (\hat{\mathsf{R}})= - \frac{R_H}{D} \left(1-\frac{1}{\hat{\mathsf{R}}}\right) \left[ \int^{\infty}_{\hat{\mathsf{R}}} \frac{\mu(\tau)}{\tau(\tau - 1)} I(\tau) \, d\tau - \frac{1}{2} \int^{\infty}_{\hat{\mathsf{R}}} \frac{I^2(\tau)}{\tau^2} \,d\tau \right].
        \end{align}
The integration of the second term by parts yields
        \begin{align} \label{mi}
            \sigma (\hat{\mathsf{R}})= \frac{R_H}{D} \left(1-\frac{1}{\hat{\mathsf{R}}}\right) \left[ \frac{I^2(\hat{\mathsf{R}})}{2\hat{\mathsf{R}}} - \int^{\infty}_{\hat{\mathsf{R}}} \frac{I(\tau)}{\tau} \left\{\frac{\mu(\tau)}{\tau - 1} - \frac{d I(\tau)}{d\tau} \right\} \, d\tau \right].
        \end{align}
Utilizing
        \begin{align}
            \frac{dI(\tau)}{d\tau} = \frac{d}{d\tau} \int^{\infty}_{\tau} \frac{\mu(\xi)}{(\xi-1)^2}\, d\xi = -\frac{\mu(\tau)}{(\tau-1)^2},
        \end{align}
it becomes evident that the second term in Equation \eqref{mi} can be calculated as
\begin{align}
            \int^{\infty}_{\hat{\mathsf{R}}} \frac{I(\tau)}{\tau} \left\{\frac{\mu(\tau)}{\tau - 1} - \frac{dI(\tau)}{d\tau} \right\} \, d\tau = \int^{\infty}_{\hat{\mathsf{R}}} \frac{\mu(\tau)}{(\tau-1)^2} I(\tau)\, d\tau = \frac{1}{2} I^2(\hat{\mathsf{R}}).
        \end{align}
Therefore, we have 
        \begin{align}
            \sigma (\hat{\mathsf{R}})= -\frac{R_H}{2D} \left(1-\frac{1}{\hat{\mathsf{R}}} \right)^2 I^2(\hat{\mathsf{R}}).
        \end{align}
We obtain the value on the shell using the condition \eqref{shellcon} as
        \begin{align}
            I^2(\hat{\mathsf{R}}_0) \simeq  \frac{\mu_0^2 \Omega^2}{(1+\Omega^2)^2 (\hat{\mathsf{R}}_0-1)^2}.
        \end{align}
Thus, the value of the second-order perturbation on the shell is
        \begin{align} \label{result2}
            \sigma(\hat{\mathsf{R}}_0) \simeq - \frac{\mu_0^2 R_H \Omega^2 }{2D(1+\Omega^2)^2 \hat{\mathsf{R}}_0^2} ,
        \end{align}
where $\hat{\mathsf{R}}_0$ can be written as a function of $\tilde{u}$ using \eqref{shellcon}. Finally, we obtain the modified $V(\tilde{u})$ function \eqref{V} owing to fluctuations by collecting the solutions \eqref{result1} and \eqref{result2}.
        \begin{align} 
            V(\tilde{u}) &=-2L \arctan \left[ \frac{R_H}{D L} \left\{ D + \ln\left(1+ e^{-\left(1-\frac{2R_H^2}{D(R^2_H+L^2)} \right)\tilde{u}}\right) \right. \right. \nonumber \\ 
            & \left. \left.\qquad \qquad \qquad \qquad \qquad \qquad  +  \frac{\mu_0 \Omega}{1+\Omega^2}\left(1+ e^{-\left(1-\frac{2R_H^2}{D(R^2_H+L^2)} \right)\tilde{u}}\right)^{-1} \right. \right. \nonumber \\
            & \left. \left.\qquad\qquad \qquad   \qquad \qquad \qquad -  \frac{\mu_0^2 \Omega^2}{2(1+\Omega^2)^2}\left(1+ e^{-\left(1-\frac{2R_H^2}{D(R^2_H+L^2)} \right)\tilde{u}}\right)^{-2} \right\}  \right].
        \end{align}
This result indicates the influence of metric fluctuations in a large $D$-dimensional AdS black hole geometry in the near-horizon regime. Interestingly, a large $D$ limit simplifies the calculations and enables us to obtain a complete solution in a compact form.

\section{Conclusions}
In this study, we investigated the effects of small fluctuations near the event horizon of high-dimensional AdS black holes. We generalized the models of \cite{york1983dynamical, barrabes1999metric, han2023metric}, where the fluctuations were approximated by the oscillations of the sources of asymptotically flat black holes, for a negative cosmological constant. In our simple model, a fluctuating black hole is described by the following higher-dimensional Vaidya-AdS solution: This results from the gravitational collapse of a massive null shell and is characterized by a mass oscillating in the spherical mode, $l=0$. To investigate the impact of small-amplitude fluctuations on the horizon and rays within black hole geometry, we employed a perturbation method to solve the null geodesic equation for radially outgoing rays. 

We found that the event horizon’s position fluctuates, with amplitudes and frequencies determined by metric fluctuation parameters. Consequently, thermodynamic variables at the horizon exhibit periodic changes, with time-averaged values incorporating variables that have correction terms proportional to the square of the amplitude parameter $\mu_0$. In particular, the modified entropy of the black hole has a slightly smaller mean value than that without fluctuations, corresponding to a stationary system. The fluctuating spacetime evolves into classical stationary spacetime following the second law $\delta S \ge 0$ of thermodynamics. Furthermore, we calculated a solution that describes the perturbed trajectory of an outgoing ray near the horizon. The impact of fluctuations can be investigated by calculating the relation $V(u)$ between the advanced time $v=V$ when the ray starts at the AdS boundary inside the null shell and the retarded time $u$ when it reaches the boundary of the black hole geometry. We derived a complete solution by introducing near-horizon coordinates defined in the large $D$ limit, which simplifies complex calculations in higher dimensions. Our results include those of \cite{han2023metric} for an asymptotically flat spacetime limit, $L \to \infty$.

We focused on spherical oscillation modes for simplicity, but exploring higher modes $l \ge 2$ could yield intriguing results. Future studies can use our results to examine both the outgoing energy flux and the asymptotic spectrum. Moreover, these findings lay the groundwork for investigating how fluctuations affect the geometry of spinning black holes.

\vspace{10pt} 

\noindent{\bf Acknowledgments}

\noindent This research was supported by Basic Science Research Program through the National Research Foundation of Korea (NRF) funded by the Ministry of Education (NRF-2022R1I1A2063176) and the Dongguk University Research Fund of 2023. BG appreciates APCTP for its hospitality during completion of this work.\\

\bibliographystyle{jhep}
\bibliography{ref_v3}

\end{document}